\documentclass{article}

     \PassOptionsToPackage{numbers, compress}{natbib}

     \usepackage[final]{neurips_2021}

\usepackage[utf8]{inputenc} 
\usepackage[T1]{fontenc}    
\usepackage{hyperref}       
\usepackage{url}            
\usepackage{booktabs}       
\usepackage{amsfonts}       
\usepackage{nicefrac}       
\usepackage{microtype}      
\usepackage{xcolor}         
\usepackage{graphicx}

\title{Centralized Intermediation in a Decentralized Web3 Economy: Value Accrual and Extraction}

\author{Dipankar Sarkar \\
  Cryptuon Research \\
  \texttt{me@dipankar.name} \\
}

\begin{document}

\maketitle

\begin{abstract}
 The advent of Web3 has ushered in a new era of decentralized digital economy, promising a shift from centralized authority to distributed, peer-to-peer interactions. However, the underlying infrastructure of this decentralized ecosystem often relies on centralized cloud providers, creating a paradoxical concentration of value and power. This paper investigates the mechanics of value accrual and extraction within the Web3 ecosystem, focusing on the roles and revenues of centralized clouds. Through an analysis of publicly available material, we elucidate the financial implications of cloud services in purportedly decentralized contexts. We further explore the individual's perspective of value creation and accumulation, examining the interplay between user participation and centralized monetization strategies. Key findings indicate that while blockchain technology has the potential to significantly reduce infrastructure costs for financial services, the current Web3 landscape is marked by a substantial reliance on cloud providers for hosting, scalability, and performance.
\end{abstract}

\section{Introduction}

The Web3 paradigm heralds an innovative phase of the Internet, focusing on decentralization, blockchain technologies, and token-based economics. This progression from Web2's centralized platforms to Web3's distributed networks proposes a transformative impact on digital interactions, asset ownership, and value creation \cite{tapscott2016blockchain}. However, the infrastructure that underpins the decentralized applications (DApps) of Web3 remains paradoxically dependent on centralized cloud services \cite{mazieres2019stellar}. This reliance presents a nuanced landscape of value accrual and power dynamics, wherein centralized entities may exert significant influence within the decentralized economy \cite{binanceresearch2021defi}.

Centralized cloud providers like Amazon Web Services (AWS), Google Cloud, and Microsoft Azure offer the computational power, storage, and networking capabilities that are integral to the functioning of blockchain networks, DeFi platforms, and NFT marketplaces \cite{zheng2018blockchain}. The role of these providers is critical, yet it seemingly contradicts the decentralization ethos of Web3 \cite{swan2015blockchain}. The interplay between the decentralized nature of blockchain-based services and the centralized characteristics of the supporting infrastructure raises important questions about the true locus of value generation and extraction within the ecosystem \cite{catalinigans2020}.

The purpose of this study is to dissect the layers of value accrual within the Web3 framework, highlighting the dichotomy between the decentralized economic activities of individuals and the centralized monetization strategies of cloud services. By scrutinizing the revenue models and the value extracted by centralized providers, and juxtaposing this with the value creation avenues available to individuals, we aim to offer a granular understanding of the economic underpinnings of Web3 \cite{weylohmmaxwell2018}.

The significance of this study lies in its potential to inform stakeholders—ranging from policymakers and developers to investors and end-users—about the dynamics of value in the Web3 ecosystem. It also aims to contribute to the discourse on how the future Internet can balance the ideals of decentralization with the practicalities of centralization, ensuring a fair and efficient digital economy \cite{buterin2014ethereum}.

In this paper, we navigate the complexities of the Web3 economy, drawing from a range of interdisciplinary sources to construct a detailed picture of where and how value is accrued and extracted. Through this exploration, we seek to offer insights into the emerging digital economy's trajectory and the potential paths it may take as it continues to evolve \cite{narayananetal2016}.

\section{Literature Review}
The literature on Web3, decentralization and the role of centralized cloud services within this emerging ecosystem is expansive and interdisciplinary, spanning technical whitepapers, economic analyses, and socio-political critiques.

\paragraph{Blockchain, the Backbone of Web3}  The foundational technology of Web3 is blockchain, a decentralized ledger that maintains a continuous and unalterable record of digital transactions \cite{nakamoto2008bitcoin}. Blockchain's potential extends beyond cryptocurrency, with implications for decentralized applications (DApps) that promise to disrupt traditional business models \cite{christidis2016blockchains}. Previous work has \cite{tapscott2016blockchain} explored the possibilities of blockchain for creating a new era of internet utility, emphasizing the importance of trust through distributed networks.

\paragraph{Centralization within Decentralization} Despite the decentralized nature of blockchain, the preponderance of centralized cloud services supporting Web3 operations has been noted \cite{swan2015blockchain}. Previous work \cite{mazieres2019stellar} argues that the current model of internet-level consensus still relies heavily on central nodes for performance and reliability. This presents a contradiction within the Web3 narrative, which aims to move away from centralized authority \cite{binanceresearch2021defi}.

\paragraph{Economics of Decentralization} Economic theories related to decentralization assess the implications of a shift from central intermediaries to peer-to-peer interactions. Previous works \cite{catalinigans2020} \cite{weylohmmaxwell2018}  delve into the economics of blockchain, offering insights into how it can lower transaction costs and enable new forms of market design; they further discuss the monopolistic tendencies that can arise even within decentralized networks, referring to them as 'decentralized monopolies'.

\paragraph{Value Creation and Extraction in Digital Economies} Research on value creation in digital economies often highlights the role of data as a key asset \cite{schilling2019technology}. In the context of Web3, this revolves around the generation, storage, and usage of data in decentralized systems. How value is extracted from these systems, particularly by centralized entities, is an area of ongoing research \cite{zohar2015bitcoin}. The role of centralized cloud providers in this value extraction process is particularly noteworthy, given their influence on the infrastructure of the Web3 ecosystem \cite{zheng2018blockchain}.

\paragraph{Value Accrual Mechanisms in Web3 Components} Various components of the Web3 ecosystem, such as DeFi, NFTs, and DAOs, provide different mechanisms for value accrual. Studies by Binance Research \cite{binanceresearch2021defi} examine how these components facilitate economic activities that challenge traditional financial systems. The literature also explores the role of oracles and smart contracts in enabling complex decentralized applications \cite{adler2018astraea}.

\paragraph{Implications for Stakeholders} The implications of the intersection between decentralized and centralized systems for various stakeholders are vast. For policymakers, the challenge lies in regulating a system that is inherently resistant to centralized control \cite{defilippi2018blockchain}. For investors and end-users, understanding where value accrues in this new economy is crucial for making informed decisions \cite{buterin2014ethereum}.

\paragraph{Future Trajectory of Web3} Finally, the literature speculates on the future trajectory of Web3. Previous works\cite{narayananetal2016} offer a comprehensive introduction to the technological underpinnings and potential future developments of blockchain technologies. They posit that the success of Web3 will depend on a balance between decentralized and centralized components, ensuring efficiency, scalability, and inclusivity.

\section{Methodology}
The methodology of this study is designed to examine the mechanisms of value accrual and extraction within the Web3 ecosystem, with a particular focus on the role of centralized cloud services. This section outlines the analytical framework, data sources, and the approach to data analysis.

\subsection{Analytical Framework}
The study adopts a mixed-methods approach, combining qualitative and quantitative analyses to explore the multifaceted dynamics of the Web3 economy. The framework is structured to evaluate the following:

\begin{enumerate}
    \item The service models and pricing strategies of centralized cloud providers.
    \item The scale of monetary transfer from decentralized to centralized entities.
\end{enumerate}

\subsection{Data Sources}
Data for this study is sourced from multiple avenues to ensure a comprehensive understanding of the Web3 economy:
\begin{itemize}
\item \textbf{Industry Reports and Whitepapers} For insights on service models, pricing strategies, and Web3 operational mechanics.
\item \textbf{Financial Statements} From leading cloud service providers to analyze revenue from Web3-related services.
\item \textbf{Academic Journals and Conferences} To incorporate theoretical and empirical findings related to Web3 economics.
\end{itemize}

\subsection{Approach to Data Analysis}
The collected data are analyzed by the following methods:

\begin{itemize}
\item \textbf{Content Analysis} Industry reports, whitepapers, and financial statements are subjected to content analysis to extract relevant information regarding business models and financial metrics.
\item \textbf{Comparative Analysis} Comparing the revenue models of centralized services with the value distribution in decentralized protocols to identify disparities and value concentrations.
\end{itemize}

\subsection{Limitations and Delimitations}
The study acknowledges the rapidly changing landscape of Web3 and the limitations in forecasting long-term trends. The analysis is also limited to the data available at the time of the study, recognizing that the opaque nature of some Web3 transactions may limit the completeness of the financial analysis.

Through this methodology, the study aims to provide an empirical foundation for understanding monetary flow and value extraction in the Web3 ecosystem, contributing to the wider discourse on the economic implications of decentralized technologies.

\begin{figure}[h]
  \centering
  \includegraphics[scale=0.4]{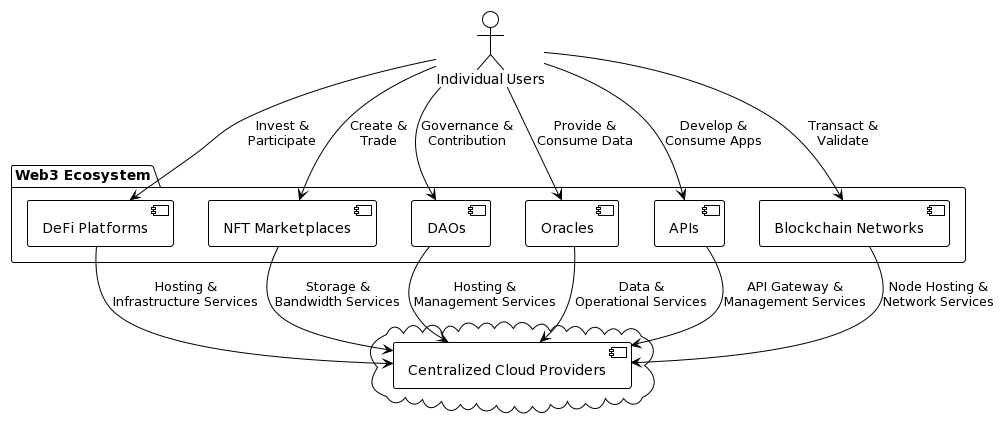}
  \caption{Value accrual across the web3 ecosystem}
\end{figure}

\section{Centralized Cloud Providers}

\paragraph{Web3 Infrastructure Hosting} Centralized cloud providers furnish the necessary infrastructure that underpins the operability of various Web3 components, offering a paradoxical foundation to the decentralized applications (DApps) they support. The infrastructure services include computational power for transaction processing, data storage for maintaining blockchain ledgers, and network bandwidth to facilitate global access and connectivity \cite{zheng2018blockchain}. 

These services are crucial for the execution and management of smart contracts, the hosting of Decentralized Finance (DeFi) platforms, the storage of Non-Fungible Tokens (NFTs), and the operation of Decentralized Autonomous Organizations (DAOs) \cite{christidis2016blockchains}.

\paragraph{Revenue Models} The revenue models adopted by centralized cloud providers are diverse, often characterized by usage-based pricing strategies that scale with the operations of their Web3 clients. These models include subscription fees for continuous access to blockchain nodes, pay-as-you-go pricing for computational resources, and tiered pricing for API calls that interact with blockchain networks \cite{mazieres2019stellar}. 

For example, cloud providers may charge DeFi platforms for the infrastructure required to handle high-frequency trading and liquidity pooling, or they may levy fees on NFT marketplaces for the storage and bandwidth used to trade and display digital collectibles \cite{binanceresearch2021defi}.

\paragraph{Examples of Engagement} Review of cloud provider engagement with Web3 entities reveal the depth of their involvement and the financial benefits they reap. 

\begin{itemize}
    \item Amazon Web Services (AWS) offers a blockchain service that supports companies like Coinbase and Gemini, which rely on AWS for their extensive cryptocurrency exchange operations \cite{aws2021managedblockchain}. 
    \item Google Cloud has partnered with blockchain protocols like Hedera Hashgraph to support their global ledger services, indicating a revenue stream from the blockchain's node operations and transaction processing \cite{googlecloud2020hedera}.
    \item Microsoft Azure's Blockchain Workbench, which facilitates the creation and management of blockchain applications, providing a suite of app development services that reduce the complexity and cost of creating blockchain applications, thus extracting value from both the developmental phase and the operational phase of Web3 projects \cite{microsoft2018azure}.
\end{itemize}

In each of these cases, the centralized cloud provider benefits financially by offering a service layer that supports the decentralized nature of the clients' projects. The engagement also shows the symbiotic relationship between Web3 entities that seek reliable, scalable infrastructure, and centralized providers that offer such services for a fee.

\section{Customer value accural}

\paragraph{Mechanisms of Earning in DeFi, NFTs, and DAOs} Individuals accrue value in the Web3 ecosystem through various innovative mechanisms, particularly within DeFi, NFTs, and DAOs. 

\begin{itemize}
    \item DeFi offers individuals opportunities such as yield farming, where users earn returns by lending crypto assets or providing liquidity to a platform \cite{schar2021defi}. 
    \item NFTs allow creators to monetize digital content directly, and collectors to potentially profit from the appreciation of digital art or collectibles \cite{nadini2021nft}. 
    \item DAOs enable individuals to participate in collective investment ventures or decentralized governance systems, often rewarding contributors with tokens that have the potential to increase in value \cite{wang2019blockchainenabled}.
\end{itemize}

\paragraph{Participation Benefits in Oracles, APIs, and Blockchain Networks} Oracles benefit individuals by allowing them to provide data to smart contracts, receiving payment for their services, which can include data from real-world events, price feeds, or other off-chain information \cite{ellis2017chainlink}. APIs provide developers with the ability to create applications that interact with blockchain networks, enabling new business models and revenue streams. By participating in blockchain networks, individuals can engage in mining or staking, earning rewards for validating transactions or maintaining the network \cite{bonneau2015sok}.

\paragraph{Examples of Value Creation} The following instances highlight the tangible benefits realized by individuals in the Web3 space. 

\begin{itemize}
    \item Artists have sold NFTs for substantial amounts, such as Beeple's digital artwork selling for \$69 million at Christie's auction \cite{christies2021beeple}. 
    \item In DeFi, platforms like Yearn.finance have turned yield farming into a lucrative opportunity for those who provide liquidity \cite{evans2020defi}. 
    \item In the context of DAOs, the DAO Maker platform allows individual investors to participate in early-stage funding rounds, democratizing access to investment opportunities \cite{daomaker2021}.
\end{itemize}

These individual success stories underscore the potential of Web3 to redistribute value creation opportunities away from traditional centralized institutions and towards individual participants. However, they also highlight the need for a deeper understanding of the risks and volatilities inherent in these nascent markets.

\section{Monetization and Value Extraction by Centralized Services}

\paragraph{Revenue Streams} Centralized cloud providers have monetized their services within the Web3 ecosystem through several key revenue streams. These include infrastructure-as-a-service (IaaS) models, providing compute power and storage; platform-as-a-service (PaaS) offerings, supplying blockchain development platforms; and software-as-a-service (SaaS) solutions, delivering turnkey applications for blockchain integration. Providers also generate revenue through data transfer fees and network usage, particularly for high-bandwidth applications such as NFT marketplaces that require significant data throughput \cite{zheng2018blockchain}.

\begin{figure}[h]
  \centering
  \includegraphics[scale=0.4]{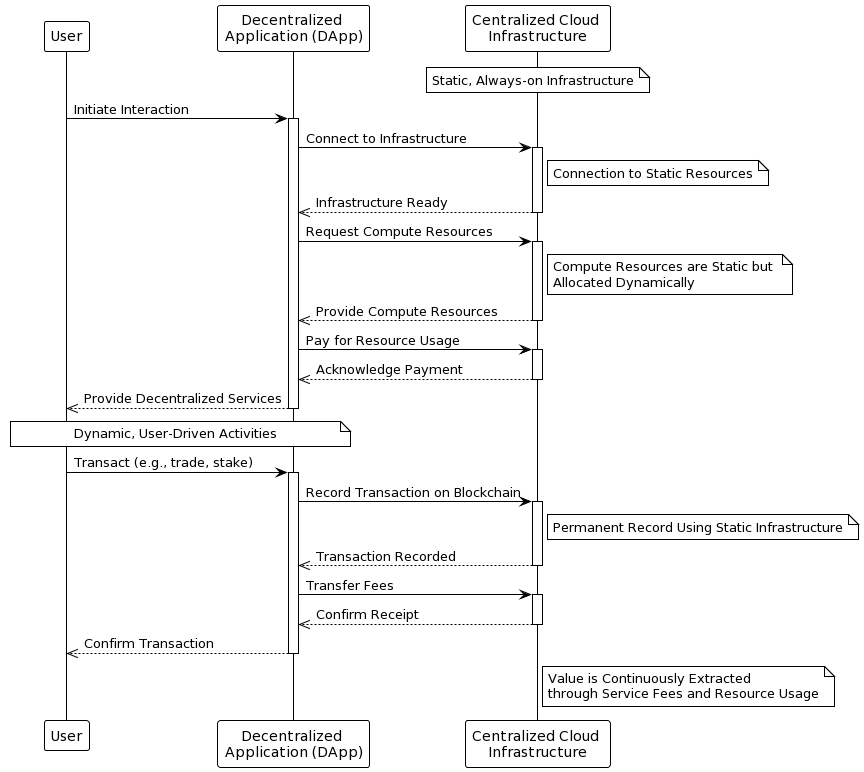}
  \caption{Example DApp Cloud interaction}
\end{figure}

\subsection{Estimation of Monetary Value}

While precise figures on the percentage of funds Web3 companies allocate to hosting services are not publicly disclosed, insights into the cost savings and efficiencies blockchain technology can introduce suggest significant interactions with centralized services. 

\begin{itemize}
    \item The Santander FinTech study, for instance, points to potential infrastructure cost reductions between US\$15 billion to \$20 billion per annum by 2022 due to blockchain integration, hinting at a substantial reallocation of financial resources that would traditionally go toward legacy systems and IT expenditures \cite{pwcblockchain}. This reallocation could imply a redirection of funds towards modern cloud hosting services equipped to handle the demands of blockchain technology.
    \item IBM's discussion on the value of blockchain networks across industries further underscores the necessity for robust and scalable infrastructure, which centralized cloud services are uniquely positioned to provide. As blockchain technology disrupts financial sectors, the dependency on cloud services for collaborative exchanges and operational capabilities is likely to grow, potentially increasing the revenue streams for these centralized providers \cite{ibmblockchain}.
    \item McKinsey \& Company's emphasis on the strategic business value of blockchain through cost reduction and operational efficiencies resonates with the monetization potential for centralized services. Given that these services underpin the operational functionality of blockchain applications, the drive for efficiency is likely to translate into sustained and potentially growing revenue for cloud providers \cite{mckinseyblockchain}.
    \item The IMF's discussion on CBDCs and the dual nature of these as both a monetary instrument and infrastructure highlights the nuanced role of centralized services in the new monetary system. As these currencies aim to improve payments by providing a public good, the infrastructure required to support them—including the services offered by centralized cloud providers—becomes increasingly critical \cite{imfcrypto}.
    \item Finally, the Bank for International Settlements (BIS) sheds light on the future monetary system, recognizing the intertwining of technological capabilities like blockchain with a superior representation of central bank money. Trust in the currency and the advantages of digital technologies imply a foundational role for secure and reliable hosting services, which can facilitate interoperability and network effects \cite{bismonetary}.
\end{itemize}

These insights collectively suggest that while the direct percentage of Web3 company expenditures on centralized hosting is challenging to pinpoint, the strategic shift toward blockchain technology in financial services is likely to drive considerable monetization opportunities for centralized cloud providers. The literature indicates a potential uptick in the value extracted by these services as the blockchain technology continues to mature and proliferate across financial sectors.

In summary, the strategic importance of blockchain technology in reducing costs and creating efficiencies in the financial sector is likely to result in a significant portion of Web3 companies' budgets being allocated to centralized cloud services. However, further research and data disclosure would be beneficial in providing a more precise understanding of the actual monetary value these services extract from the Web3 ecosystem.

\subsection{Comparative Analysis with Traditional Digital Economies}

Comparing the Web3 economy with traditional digital economies reveals a shift in value extraction methods. In traditional digital economies, centralized platforms extract value through advertising, data monetization, and subscription models. In contrast, within Web3, the value is extracted through service provision to a decentralized ecosystem, which includes transaction processing, smart contract deployment, and node hosting services. This shift represents a move from data and attention monetization to the monetization of decentralized infrastructure and services \cite{tapscott2016blockchain}.

Despite the growing significance of blockchain-based services, the overall revenue from Web3 might still be a fraction of what is generated by traditional centralized platforms, indicating the nascent stage of the Web3 economy. The comparison highlights the potential for growth in Web3 and the opportunity for centralized services to expand their revenue models as the technology and its adoption mature.

\subsection{Gaps and Future Research Directions}
The primary gap in this analysis is the lack of transparent and accessible financial data directly attributing cloud provider revenues to Web3 services. The estimation of value extraction would benefit from more granular data disclosures by cloud service providers regarding their blockchain and Web3 customer segments. Additionally, the rapid pace of change in Web3 technologies and economic models necessitates ongoing data collection and analysis to keep estimations current.

Further research should aim to map the evolution of revenue models as the Web3 economy matures, assess the impact of technological advancements on cost structures, and explore the potential for new monetization strategies beyond the current service-based models. Longitudinal studies would be particularly valuable in tracking these trends over time.

\section{Discussion}

The increasing reliance of decentralized applications on centralized cloud services has notable implications for the Web3 economy. The Santander FinTech study underscores a potential reduction in infrastructure costs due to blockchain, which could lead to a significant monetary shift from traditional IT spending to cloud services \cite{pwcblockchain}. This reallocation highlights a centralization of monetary flow to cloud providers who hold the infrastructural keys to the decentralized world, raising questions about the concentration of power and potential single points of failure \cite{tapscott2016blockchain}.

\paragraph{Potential Risks and Benefits} For stakeholders, such as Web3 companies, there is a dual-edged sword. On one hand, leveraging centralized cloud services can offer scalability, reliability, and advanced services that are not yet feasible in a fully decentralized environment \cite{ibmblockchain}. On the other hand, dependency on these services introduces risks such as increased costs, potential data privacy issues, and the risk of service discontinuity due to centralized control \cite{zheng2018blockchain}.

End-users and consumers may benefit from the enhanced performance and stability that centralized services provide to the decentralized applications they use. However, they also face the risk of censorship and the possibility of compromised data integrity if these central providers become malevolent actors or are compromised \cite{mckinseyblockchain}.

\paragraph{The Dichotomy} The inherent tension between the decentralized ethos of Web3 and the centralized nature of its current infrastructure is a fundamental dichotomy that underpins the Web3 space. While decentralization promises to distribute power among users and remove intermediaries, centralization—particularly in value accrual—suggests a concentration of economic control and influence \cite{bismonetary}.

The IMF's discussion on CBDCs illustrates how new forms of currency could balance this dichotomy by offering a decentralized monetary instrument built on centralized infrastructure, potentially providing a public good while maintaining the trust and stability associated with central bank oversight \cite{imfcrypto}.

The BIS adds to the conversation by suggesting that technological capabilities like blockchain could be integrated with central bank money, creating a hybrid system that benefits from the innovations of decentralization while maintaining the trust and authority of centralized institutions \cite{bismonetary}.

\paragraph{Summary} The Web3 ecosystem is navigating a complex interplay between the decentralization of operations and the centralization of infrastructure. This dynamic presents both opportunities and challenges for all stakeholders involved. It calls for a careful evaluation of how to harness the benefits of both worlds to create a resilient, efficient, and inclusive digital economy.

\section{Conclusion}

This paper has explored the complex dynamics of value accrual and extraction within the Web3 ecosystem, particularly the paradoxical reliance on centralized cloud providers for the deployment of decentralized applications. Key findings indicate that while blockchain technology has the potential to significantly reduce infrastructure costs for financial services \cite{pwcblockchain}, the current Web3 landscape is marked by a substantial reliance on centralized services for hosting, scalability, and performance \cite{ibmblockchain}. This dependency has introduced a centralization of economic control, potentially conflicting with the decentralized ethos of Web3 \cite{tapscott2016blockchain}.

\subsection{Future Outlook} 

Looking forward, the Web3 economy is poised at a crossroads between further entrenchment of centralized services and a move towards genuine decentralization. As technologies mature, there is potential for a more balanced value distribution that leverages the benefits of centralization—such as reliability and efficiency—while cultivating the growth of decentralized alternatives that could mitigate risks associated with central points of control and failure \cite{bismonetary}. The evolution of CBDCs could exemplify this balance, marrying the stability of traditional financial systems with the innovation of blockchain technology \cite{imfcrypto}.

\subsection{Recommendations}

For policymakers, the recommendation is to foster a regulatory environment that encourages innovation in decentralized technologies while ensuring robust consumer protections and data privacy standards. As the BIS suggests, trust in digital currencies—whether decentralized or centralized—remains paramount, necessitating clear guidelines and oversight \cite{bismonetary}.

Developers are encouraged to innovate towards reducing reliance on centralized infrastructures, perhaps by building more resilient decentralized hosting solutions and exploring hybrid systems that can offer decentralization at scale \cite{zheng2018blockchain}.

Participants in the Web3 space should remain vigilant about the concentration of control and the implications it has for the security and integrity of the decentralized services they use. They should advocate for transparency and support initiatives that aim to distribute value and control more evenly across the ecosystem.

\subsection{Summary}

While the current state of Web3 indicates a significant centralization of infrastructure, there is a clear pathway towards a more decentralized future. By working collaboratively, stakeholders across the spectrum can contribute to an economic model that aligns more closely with the original principles of Web3, fostering an ecosystem that is not only innovative and efficient but also equitable and resilient.

\bibliographystyle{abbrvnat}
\bibliography{sample}
\end{document}